\documentstyle[preprint,eqsecnum,aps]{revtex} 
\tighten
\begin{document}
\def\cf{{\it cf.\ }}
\def\eff{{\rm eff}}
\def\ed{{\sl ed.~by\ }}
\def\eg{{\it e.g.,\ }}
\def\Eg{{\it E.g.,\ }}
\def\ie{{\it i.e.,\ }}
\def\Ie{{\it I.e.,\ }}
\def\etal{{\it et al.~}}
\def\etc{{\it etc.}}
\def\via{{\it via}}
\def\viz{{\it viz.~}}
\def\Eq#1{Eq.~(#1)}     
\def\eq#1{Eq.~(#1)}     
\def\Eqs#1{Eqs.~(#1)} 
\def\eqs#1{Eqs.~(#1)} 
\def\frac#1#2{{\textstyle #1 \over \textstyle #2}} \def\half{{\textstyle {1
\over 2}}}
\gdef\journal#1, #2, #3, 1#4#5#6{       
{\sl #1~}{\bf #2}, #3, (1#4#5#6)}       
\draft
\date{\today}

\title{Equivalent Sets of Histories \\ and Multiple Quasiclassical Realms}

\author{Murray Gell-Mann  and James B.
Hartle\thanks{\narrowtext Permanent address: Department of Physics,
University of
California, Santa Barbara, CA 93106-9530, hartle@@cosmic.physics.ucsb.edu
\widetext}}
\address{Theoretical Astrophysics Group (T-6)\\ Los Alamos
National Laboratory, Los Alamos, New Mexico 87545\\ and\\ Santa Fe
Institute, 1399 Hyde Park Road, Santa Fe, New Mexico 87501\\ and\\
Center for Advanced Studies,\\
University of New Mexico, Albuquerque, New Mexico 87131\\ and\\
Isaac Newton Institute for Mathematical Sciences\\
University of Cambridge, Cambridge, CB3 0EH, UK}

\maketitle

\begin{abstract}

We consider notions of physical equivalence of sets of histories in the
quantum mechanics of a closed system
such as the universe. We show first how the same set of histories can be
relabeled in various ways, including the use of the Heisenberg equations
of motion and of alias (passive) transformations of field variables.
In the contrasting case of the usual approximate quantum mechanics of a
measured subsystem of the universe, two
observables represented by different Hermitian operators (as opposed to the
same operator relabeled) are physically
distinguished by the different pieces of apparatus used to measure them.
That is true even
if they are related by a unitary transformation  and  the
state of the system is such that the probabilities of ranges of values of
the observables are the same. In the quantum mechanics of a closed system,
however, any apparatus is part of the system
and the notion of physically distinct situations has a different character.
 Making our previous
suggestions more precise, we show that a triple consisting of an
initial condition, a Hamiltonian, and
a set of histories is physically equivalent to another triple
if the operators representing these initial conditions, Hamiltonians,  and
histories are related by any fixed unitary transformation.
We apply this result to the question of whether the
universe might exhibit physically inequivalent quasiclassical realms
(which we earlier called quasiclassical domains), not
just the one that includes familiar experience.  We describe, in more detail
than we
have before, how the probabilities of alternative forms, behaviors, and
evolutionary histories of information gathering and utilizing systems (IGUSes)
using the usual quasiclassical realm could in principle be calculated
in quantum cosmology, although it is, of course, impractical to perform
the
computations. We discuss how, in principle, the probabilities of occurence
of IGUSes
could be calculated in realms distinct from the usual quasiclassical
one --- realms such as Lloyd's representation of the universe as a quantum
computer. We discuss how
IGUSes       adapted mainly to two different realms could draw
inferences about each other using a hybrid realm consisting
of alternatives drawn from each.

\end{abstract}

\pacs{}

\narrowtext
\setcounter{footnote}{0}
\section{Introduction}
\label{sec:intro}

Quantum mechanics, in its most general form, predicts the probabilities
of alternative decohering coarse-grained histories of the universe. The
computation of these probabilities requires an initial condition, given
by a density matrix $\rho$, and a narrative description of the histories
expressed in terms of suitable operators. Suppose, for simplicity that
the density matrix is pure. What gives rise to
probabilities in quantum mechanics is the mismatch between the state
vector of the universe and the state vectors, orthogonal to each other,
associated with the individual decoherent histories in a set of
alternative histories. When the state vector of the universe is resolved
into a sum of vectors corresponding to the histories in the set, the
norms of those vectors are the probabilities. The specification of a set
of alternative decoherent histories is just as important for the
utilization of quantum mechanics as the characterization of the
Heisenberg state vector of the universe (or the equivalent Schr\"odinger
state vector and its time development). It is not just the state of the
universe that gives quantum mechanics its meaning. It also matters
which set of questions is asked of that state.

Among all the possible sets of alternative histories for which
probabilities are predicted by the quantum mechanics of the universe, those
describing a quasiclassical realm\footnote{In our previous work we have
referred to ``quasiclassical domains''. We now suggest using instead
the expression ``quasiclassical realms'' to avoid confusion with the
usual meaning of ``domains'' in physics. We use the word ``realm'' as
a synonym for ``decohering set of alternative histories''.}, like the one that
includes familiar
experience,
are of special importance. By a quasiclassical realm we mean roughly a set
of histories (or a class of nearly equivalent sets) maximally refined,
consistent with obeying a realistic principle of decoherence and with
exhibiting
patterns of approximately deterministic correlations governed by
phenomenological classical laws connecting similar operators at
different times \cite{GH90a}.
(Those patterns are interrupted, of course, by frequent
small fluctuations and occasional major branchings of histories.) Such
quasiclassical realms are
important for at least two reasons: (1) The existence of at least one
quasiclassical realm appears to be a reasonable extrapolation from
empirical fact and should therefore be a prediction in quantum cosmology
from the
fundamental theory of the elementary particles and the initial condition of the
expanding universe. That usual quasiclassical
realm is defined by alternative ranges of values
of certain operators (called usual quasiclassical operators), which are
particular
kinds of local operators (such as electromagnetic fields or densities of
conserved or nearly conserved quantities)  averaged over small regions of
space throughout the universe, at a sequence of times that span the whole
history of the universe. (2) Coarse grainings of this usual quasiclassical
realm
are what we (humans and many other systems) use in the process of gathering
information about the universe and making predictions about its future. We
deal directly with values of usual quasiclassical operators, to some of
which our
senses are adapted. Other quantum-mechanical operators are accessible when
correlated with these quasiclassical ones, i.e., in measurement situations.

There are very many of sets of histories that decohere, and
trivial examples of exactly decohering sets can easily be exhibited
that are nothing like a quasiclassical realm, let alone the quasiclassical
realm that includes everyday experience
\cite{GH90a,DKup}. To understand what quasiclassical realms are possible in
quantum mechanics, it is desirable to make more mathematically precise
the measure of classicality on the collection of
all sets of decohering histories. The refinement of the
definition \cite{GH90a,GH93a,PZ93} can help to answer the difficult and
fundamental question of whether the quantum mechanics of the
universe exhibits only an essentially unique
quasiclassical realm or whether there are essentially different
ones. It can help to give a general characterization of quasiclassical
operators, the values of which specify the alternative histories of a
quasiclassical realm and to, in principle,  derive the form of the
phenomenological deterministic laws that approximately govern a given
quasiclassical realm as has already been done in some model problems
\cite{GH93a}. (That form  tends to be far removed from the form of the
equations describing the underlying fundamental dynamics, \eg heterotic
superstring theory.) In connection with the phenomenological laws, it
is important analyze
the coarse graining necessary
to achieve approximate classical predictability in the
presence of the noise that typical mechanisms of decoherence produce.
Finally, the coarse graining(s)
used to define entropy in the second law of thermodynamics should be
connected the coarse
graining used to define the histories of a quasiclassical realm \cite{GH95}.

We shall not review here the efforts to achieve all of these
objectives nor shall we investigate the details of the measure.
Rather, we discuss, as a prerequisite, the nature of
physical equivalence between sets of
coarse-grained histories of a closed system, so as to understand better what
it would mean for the universe to exhibit essentially inequivalent
quasiclassical realms.  We then examine some implications for information
gathering and utilizing systems (IGUSes) of such inequivalent realms
(and of certain other realms as well).

We mentioned in \cite{GH90a} that the notion of physically distinct sets of
alternative histories has a different character for a closed
quantum-mechanical system and for the approximate quantum mechanics of
measured
subsystems. We make this difference precise in Section II of this article.
We first show how the description of a given set of histories, constructed from
alternatives at a sequence of times, may be varied in several different ways
without affecting the histories themselves.
First, making use of the Heisenberg equations of motion to change the
description of the alternatives in terms of fundamental fields, we can
reassign the times of the alternatives (as long as their time-order is
maintained.) Second, the alternatives may be relabeled by making alias
(passive) transformations of the fundamental fields and conjugate
momenta.

Under a relabeling of the above types, operators representing the
histories remain unchanged. However, in the quantum mechanics of a closed
system described in terms of quantum fields, even sets of histories
represented by {\it different} operators
may be physically equivalent.
That is not true
in the usual approximate quantum mechanics of measured subsystems, where
two observables
represented by different Hermitian operators are physically distinguished
by the different kinds of apparatus (outside the subsystem) used to measure
them. In the quantum mechanics
of a closed system, however, any apparatus is part of the system and triples
of Hamiltonians, initial conditions, and sets of histories
represented by different operators may be physically indistinguishable.
Suppose that a fixed unitary transformation acts on the Hamiltonian, the
density
matrix representing an initial condition, and on the projection operators
representing alternatives at moments of time in a set of histories, but not
on the fundamental
fields in terms of which all these operators are described. We show
in Section II that the resulting new initial conditions and sets of
histories are physically equivalent to the old ones in the sense that they
admit an identical description in terms of fundamental fields, with the
same probabilities for corresponding histories. Initial conditions and sets
of histories related in this way should be identified with each other and
placed together in physical equivalence classes.\footnote{Similar notions
of physical equivalence will be discussed in \cite{DKup}.}

This relationship of physical equivalence is important for the problem of
quasiclassical realms. Measures of classicality should be on physical
equivalence classes. Given an initial density matrix $\rho$ and a set of
histories constituting a quasiclassical realm,  we do not exhibit another
distinct quasiclassical realm by transforming the projection operators
of the first realm using a constant unitary transformation that leaves $\rho$
invariant.\footnote{The question of the relationship between sets of
histories related by unitary transformations that leave the initial density
matrix invariant was first raised for one of the authors (JBH) in
discussions with R.~Penrose in 1989 where the question of whether they
should be identified also arose.}
The two sets are physically
indistinguishable.

As observers of the universe, we (and all other IGUSes that we know of) make
use of a particular quasiclassical realm (further
coarse grained according to the limitations of our senses and instruments.)
The explanation for this is
not to be sought in some privilege conferred on quasiclassical realms by
the quantum mechanics of closed systems, for quasiclassical realms are but
a small subset of the collection of all sets of decoherent histories\footnote{
Cf. the remarks in \cite{Har95}.}, and moreover IGUSes,
including human beings, occupy no such special place and play no such
preferred role
in this formulation of quantum mechanics as they do in the ``Copenhagen''
interpretation(s)	. Rather, as we suggested in
\cite{GH90a}, it is plausible, in the context  of quantum cosmology, that
IGUSes evolve by exploiting realms with a high level of predictability
such as quasiclassical realms, focusing on variables
that present enough regularity over
time to permit the generation of models (schemata) with significant
predictive power.

It is, of course, an impractical task to
compute the probabilities of alternative evolutionary tracks for IGUSes
from the fundamental quantum-mechanical theory of the elementary particles
and of the initial condition of the universe. Nevertheless, it is
clarifying to investigate how such questions might be posed in principle in
quantum
cosmology even if we can only guess the answers. We offer some thoughts on
this topic in Section III.

If only one collection of essentially equivalent sets of decoherent
histories with high classicality emerges from the initial condition of
the universe and the dynamics of the elementary particles,
then the usual quasiclassical realm is essentially unique.
However, if the quantum mechanics of the universe exhibits essentially
inequivalent
quasiclassical realms then it is possible that IGUSes evolve on
branches of more than one of them. Moreover, there may be other realms,
even more
deterministic, in which IGUSes arise, for example, a realm in which the
universe behaves as a quantum computer \cite{Llo93}.
We discuss these matters in Sections III, IV, and V.

\section{Physically Equivalent Sets of Histories} \label{sec:II}
In this Section we shall describe a notion of physical equivalence
between sets of alternative histories of a closed quantum-mechanical
system, in the general formulation of quantum mechanics appropriate
to such systems.  Many readers will be more familiar with the
``Copenhagen'' formulation of quantum mechanics, usually explained in
textbooks,
 which is concerned with
predicting the outcomes of measurements on a subsystem. These two
formulations are not in conflict with each another.
The usual ``Copenhagen'' formulation is an {\it approximation} to the
more general quantum mechanics of a closed system and is applicable
to sets of histories describing  measurement situations when
certain approximate features of these histories
can be idealized as exact \cite{Har91a}. We begin by describing
physical equivalence in the quantum mechanics of a closed
system and return in Section I to the more restrictive notion valid in
the approximate quantum mechanics
of measured subsystems.

\subsection{The Quantum Mechanics of a Closed System} \label{sec:a}

To establish some notation and clarify our assumptions, we give a brief
review of the quantum mechanics of a closed system.\footnote{We follow our
earlier work, for example, \cite{GH90a}, \cite{Har91a}, \cite{GH93a}.} We
consider such a system, most generally and accurately the universe as a
whole, including both observers and observed, both measuring apparatus and
measured
subsystems, if any. We work in the approximation in which the geometry of
spacetime is approximately fixed and gross fluctuations in it
are neglected.\footnote{It is a special virtue of our approach that these
restrictions are not actually necessary.  For a generalized quantum theory that
 can
incorporate dynamical spacetime geometry see \cite{Harpp} and earlier
references therein.} We also assume that spacetime is foliable by spacelike
surfaces. Times are then well-defined and the usual formalism of Hilbert
space, states, Hamiltonian, etc. can be used to describe quantum theory. We
assume a fundamental quantum field theory. We shall usually indicate just a
single scalar field $\phi(x)$, hoping that the reader may
make the straightforward generalization to the usual panoply of Fermi,
tensor, and
other fields (or to superstring theory, in which there is something like an
infinite set of such fields). The dynamical evolution of the field through a
family of spacelike surfaces is generated by a Hamiltonian which, on a
spacelike surface labeled by $t$, is a functional of the field on that
surface $\phi({\bf x}, t)$ and its conjugate momentum $\pi({\bf x}, t)$.
The canonically conjugate pair satisfy the fundamental commutation
relations \begin{equation} \bigl[\phi({\bf x}, t), \pi({\bf x'}, t) \bigr] =
i \delta ({\bf x}, {\bf x}^\prime),
\label{twofour}
\end{equation}
where $\delta({\bf x}, {\bf x}^\prime)$ is the $\delta$-function on the
spacelike surface. We use units for which $\hbar=1$.

Various quantities represented by Hermitian operators ${\cal O}[\phi({\bf
x}, t), \pi({\bf x}, t)]$ can be constructed from the fields and momenta on
a spacelike surface. Projections onto an exhaustive set of alternative
ranges of these quantities define alternatives at the moment of time $t$.
Giving a sequence of such sets of alternatives at times $t_1, \cdots, t_n$
defines a
set of alternative histories for the closed system, although not of
the most general kind, as we shall see. We denote the
sets of projections by $\{P^k_{\alpha_k} (t_k)\}$, $k=1, \cdots, n$
where the superscript $k$ distinguishes the quantity ${\cal O}$ and the set
of ranges employed at time $t_k$, and $\alpha_k$ the particular
range represented by the projection. The operators $\{P^k_{\alpha_k} (t_k)\}$
satisfy
\begin{equation}
P^k_{\alpha_k} (t_k) P^k_{\alpha^\prime_k} (t_k) =
\delta_{\alpha_k\alpha^\prime_k} P^k_{\alpha_k} (t_k)\ , \qquad
\Sigma_{\alpha_k} P^k_{\alpha_k} (t_k) = I, \label{twofive} \end{equation}
showing that they are projections representing an exhaustive
set of mutually exclusive alternatives.

The Heisenberg equations of motion
\begin{mathletters}
\label{eq:twoseven}
\begin{eqnarray}
\phi({\bf x}, t^\prime)& = &e^{iH(t^\prime-t)} \phi({\bf x}, t) \ e^{-
iH(t^\prime-t)},\label{eq:aaa}\\
\pi({\bf x}, t^\prime)& = & e^{iH(t^\prime-t)} \pi({\bf x}, t) \
e^{-iH(t^\prime-t)},\label{eq:bbb}
\end{eqnarray}
\end{mathletters}
allow an operator ${\cal O}[\phi({\bf x}, t), \pi({\bf x}, t)]$ to be
reexpressed in terms of the field and momentum  at another time. Thus in the
Heisenberg picture each exhaustive set of orthogonal projection operators
may be regarded, for any time,  as a set of projections on ranges of some
quantity
at that time.
Given a set of projections satisfying (\ref{twofive}) , an arbitrary time
can be assigned, and the
projections expressed in a suitable manner in terms of the field and momentum
operators at that time.\footnote{
Alternatives lacking a well defined time,
such as averages of fields over ranges of time, are not represented
in terms of projections
onto ranges of the corresponding operators. To do so would generally leave
the time ordering of such projections ambiguous. Rather, such spacetime
alternatives are represented by {\it sums} of possibly continuous chains of
projections.
For fuller
details see \cite{Har91b}, \cite{Harpp}.}

Each sequence of alternatives $(\alpha_1, \cdots, \alpha_n)$ at definite
moments of time defines one member of
a set of possible alternative histories of the closed system.
Such histories are represented by the corresponding time-ordered chains of
projection operators. A completely fine-grained set of histories would be
defined by sets of one-dimensional projections (projections onto a basis
for Hilbert space) at each and every time. There are infinitely many
different sets of
fine-grained histories corresponding to the different choices of basis at
each time.\footnote{The problem of the physical equivalence of sets of
histories would be considerably simplified if there were a unique
allowed set of fine-grained histories. A particular distinguished set, paths in
the configuration space of quantum fields, is the starting point for a
standard sum-over-histories
formulation of quantum mechanics. However, for generality, we allow here
all the other fine-grained histories that can be constructed by transformation
theory.} The most general notion of a set of alternative histories is a
partition of one of these sets of fine-grained histories into exclusive
classes $\{c_{\alpha}\}$. The individual classes are the individual
coarse-grained histories and are represented by
class operators, $C_\alpha$, which are sums of chains of the
corresponding projections in the class:
\begin{equation} C_\alpha =
\sum\limits_{(\alpha_1, \cdots, \alpha_n)\epsilon\alpha} P^n_{\alpha_n}
(t_n) \cdots P^1_{\alpha_1} (t_1) \label{twoeight} \end{equation}
where we allow the possibility of an infinite sequence of times. Although
we have not indicated it explicitly, such sets of histories are generally
branch-dependent --- the set of operators $\{P^k_{\alpha_k} (t_k)\}$
may depend on the previous specific alternatives $\alpha_1, \cdots,
\alpha_{k-1}$ and times $t_1, \cdots, t_{k-1}$ and should really be written
$\{P^k_{\alpha_k}(t_k;\alpha_{k-1},t_{k-1},\alpha_{k-2},t_{k-2},\cdots,
\alpha_1,t_1)\}$.
(We have assumed causality, and so only previous alternatives matter.)

Probabilities are predicted for members of a set of alternative histories
of a closed system when there is negligible quantum-mechanical interference
between them. Interference between a pair of histories is measured by the
decoherence functional \begin{equation}
D\left(\alpha^\prime, \alpha\right) = Tr\left(C_{\alpha'} \rho
C^\dagger_\alpha\right)\ ,
\label{twonine}
\end{equation}
where $\rho$ is a density matrix representing the initial condition of the
closed system.\footnote{We thereby restrict ourselves to the usual quantum
cosmology in which a distinction is made between the past, with an initial
condition represented by $\rho$, and the future, with a condition of
effective indifference with respect to final state. Similar notions of
physical equivalence can be introduced in the time-neutral generalizations
of the usual quantum mechanics of closed systems (with initial and final
conditions)
that have been discussed (\eg in \cite{Gri84}, \cite{GH93b}), largely as
``straw man'' theories.} When the ``off-diagonal''
elements of $D$ are sufficiently small the set of histories is said to
(medium) decohere;\footnote{We use medium decoherence for illustrative
purposes. Our considerations would also apply to weak decoherence and to
the still weaker consistency conditions of Griffiths \cite{Gri84} and
Omn\`es \cite{Omnsum}. However, in realistic cases medium decoherence
obtains, or an even stronger condition.} the diagonal elements are then the
probabilities
$p(\alpha)$ of the individual histories in the set; \viz \begin{equation}
D\left(\alpha^\prime, \alpha\right) \approx \delta_{\alpha^\prime\alpha} p
(\alpha)\ . \label{twoten}
\end{equation}
In particular, when the initial $\rho$ is pure, $\rho=|\Psi\rangle
\langle\Psi|$, and the histories of the set are all (medium)
decoherent, one has
\begin{equation}
p(\alpha) = \left\| C_\alpha |\Psi \rangle \right\|^2\ . \label{twoeleven}
\end{equation}
\subsection{The Description of Histories and the Trivial Relabeling of
Hilbert Space} \label{sec:b}

As the preceding discussion shows, the objects of interest in quantum
theory are triples $(\{C_\alpha\}, H, \rho)$ consisting of operators
$\{C_\alpha\}$ of the form (\ref{twoeight}) representing a set of
alternative coarse-grained histories of the closed system, a Hamiltonian
$H$ connecting
field operators at different times through Heisenberg equations of
motion, and a density matrix $\rho$ representing the initial condition.
Given a Hilbert space $\cal H$, it is possible in principle to enumerate
mathematically, without reference to the fundamental fields,
 all the operator triples that represent decohering sets
of histories, Hamiltonians, and initial conditions.
However, as we stressed in \cite{GH90a}, `` it is clear that the
mathematical problem of enumerating the sets of decohering histories in
a given Hilbert space has no physical content by itself. No description
of the histories has been given.  ...
 No distinction has been made between
one vector in Hilbert space as a theory of the
initial condition and any other. The resulting probabilities, which
can be calculated, are merely abstract numbers.''

As we further discussed in \cite{GH90a}, the sets of possible triples
acquire physical content when the operators corresponding to the
fundamental fields $\phi({\bf x},t)$ are specified in $\cal H$ so that,
for example, the eigenvectors of the smeared field operators are fixed.
Any set of orthogonal projections
$\{P_\alpha(t)\}$ at time $t$ can be described as projections onto ranges
of some
operator ${\cal O}[\phi({\bf x},t), \pi({\bf x},t)]$ at that time.
It is then possible to give a narrative describing each member of a
set of alternative histories $\{C_\alpha\}$. Contact with the fundamental
interactions is made when the Hamiltonian $H$ is expressed in terms of
the field operators. Initial conditions are distinguished when $\rho$ is
described in terms of fields. The probabilities thereby acquire
physical meaning as the probabilities of alternative histories of the
universe, with a particular Hamiltonian and initial
condition.

There is some arbitrariness in the choice of subspaces of the mathematical
Hilbert space identified as having definite values of the smeared
fields. This identification can be changed by transforming {\it all}
operators in the theory --- the $C_\alpha$'s, $H$, $\rho$, {\it and}
the fields $\phi(x)$ --- by a fixed unitary transformation. The
result is merely a relabeling of Hilbert space with no physical
consequences. Sets of triples and fields so related are clearly
physically equivalent. Only an alias (passive) transformation has
been carried out.

While the quantum mechanics of closed systems clearly exhibits this
trivial notion of physical equivalence, it also exhibits further, less
trivial kinds of physical equivalence arising from the Heisenberg
equations of motion and from the invariance of the theory under field
redefinitions. In the remainder of this Section we explore those kinds of
equivalence.

\subsection{The Same Operators Described in Terms of Fields at Different Times}
\label{sec:c}

We mentioned
earlier that in the Heisenberg picture a given projection operator could be
assigned to an arbitrary time by using the equations of motion to determine
its form in terms of the field operators.
For example, in the case of a free particle with mass $m$ moving in
one dimension, the Heisenberg equations of motion have the solutions
\begin{mathletters}
\label{eq:twothirteen}
\begin{eqnarray}
x(t) & = & x(0) + p(0) t/m\ ,
\label{eq:a1}\\
p(t) & = & p(0) \ .
\label{eq:b2}
\end{eqnarray}
\end{mathletters}
Thus, a projection onto a range $\Delta$ of the position operator $x(6)$
for time $t=6$ might equally well be described as the projection onto
the range $\Delta$ of the operator $x(0)+p(0)6/m$ referring to $t=0$.
These projection operators are equal as a consequence of the equations of
motion.\footnote{
The analogous situation in classical physics may be clarifying.
In the classical ``Heisenberg picture,'' in which the alternatives are
regions of phase space that vary in time according to the equations of
motion,
the region of phase space in which
the initial $(x(0), p(0))$ are such that $x(0) + p(0)t/m$ lies in a range
$\Delta$ is the {\it same} as the region in which $x(t)$ lies in $\Delta$,
as a consequence of the time evolution (\ref{eq:twothirteen}). }

For a closed system the
equality of such operators assigned to different times, arising from
the Heisenberg equations
of motion,
is absolute, since there is no way to alter these Heisenberg equations
by means of an external perturbation. Thus
a set of projection operators described in terms of fields at two
different times does not correspond to two different sets of
alternatives for the closed system, but rather to the same set of alternatives
described in two different ways. Put differently, in a closed system
there are no external clocks to give an independent meaning to a moment
of time.\footnote{ In
general relativity, which is invariant under reparametrizations of the time
co\"ordinate, the notion of alternatives at a moment of time requires
careful examination, and in quantum gravity, where there is no fixed
background spacetime geometry, we cannot expect to have the same notion
of time as is described in this subsection, where spacetime geometry
has been assumed fixed.  For an approach to a generalized quantum mechanics
of spacetime geometry see \cite{Harpp}.}
Of course, there may be good reason for $\it us$ to prefer one
description to another. In particular, the lengths of the
two descriptions may be different, but there is no physical
distinction between them.

Thus, in a set of alternative histories consisting of sequences of sets
of alternatives at definite moments of time [{\it cf.}(\ref{twoeight}],
the times of the alternatives may be assigned arbitrarily although it is
convenient -- and necessary to avoid ambiguity --
to keep their order the same as the order of the sets so that
the histories are narratives proceeding forward in time.
The order of the projection operators is important,
for different sets of projection operators do not necessarily commute.
Since a change in the the values of the times merely corresponds to a
different description of
the histories, the decoherence and probabilities of the set of histories
must be unaffected by such a change in description. In fact,
the operators themselves are unchanged
and hence the decoherence functional remains the same.

\subsection{Fields as Co\"ordinates}             \label{sec:d}

Fields and their conjugate momenta are canonical co\"ordinates on the
phase space of classical field theory, and classical field redefinitions
mediate between different choices of these
canonical co\"ordinates. Classical mechanics may be formulated in a generally
covariant manner, allowing arbitrary choices of canonical co\"ordinates.
In a similar way, the quantum mechanics we have been using allows
arbitrary choices of the canonical pairs of field operators and their
conjugate momenta satisfying (\ref{twofour}).

In classical mechanics it is possible to fix a co\"ordinate system on
phase space by requiring a sufficient number of physical quantities
to have specified functional forms in terms of the co\"ordinates. For
example, in the classical mechanics of a system of particles, one could
require that the Cartesian co\"ordinates of the displacement of each
particle from a fixed origin be equal to a co\"ordinate set $\{q^i\}$.
Similarly, in  quantum theory one could presumably eliminate the freedom
to make field redefinitions by requiring certain physical quantities
({\it e.g.} the Hamiltonian, momentum, ... {\it etc.}) to have
definite functional forms in terms of the field variables.
The subspaces of Hilbert space are then labeled by particular
physical quantities. As a result no issue of equivalent descriptions
through field redefinitions arises --- a particular kind of description
has been singled out by convention. Even in that case, of course, the
freedom remains to make transformations corresponding to exact
symmetries of the Hamiltonian.

Labeling the Hilbert space by physical quantities is a convenient
approach to the quantum mechanics of simple, particular systems where
small numbers of physical quantities are readily identified, as in
discussions in typical textbooks. Some authors prefer to assume that in
{\it any } discussion Hilbert space has been implicitly labeled by
physical quantities. That is possible, but inconvenient for
discussing quantum mechanics in general rather than merely applying
it to some specific system. For that reason we prefer to discuss the
general form of the theory, where field redefinitions are not excluded
by convention and different descriptions of the same physical situation
are possible in terms of different fields. We leave the vectors of Hilbert
space
unlabeled until labels are explicitly assigned.  Our subsequent discussion of
physical equivalence should be understood in this context.
\hfill\break

\subsection{Different Descriptions of the Same Histories Through
Field Redefinitions}
\label{sec:e}
A consequence of the general treatment of fields outlined
above is that histories can be described in different ways through
field redefinitions.
We discussed above how projection operators representing alternatives at a
moment of time
could be described as projections onto ranges of values of operator
functions of fundamental fields and their conjugate momenta. The Hamiltonian
$H$ and the density
matrix $\rho$ representing the initial condition may be similarly
described. However, given one conjugate pair $\bigl(\phi(x), \pi(x)\bigr)$
satisfying (\ref{twofour}), it is possible to find other canonical pairs
through field redefinitions
\begin{equation}
\widetilde\phi(x) = \widetilde\phi (x ; \phi(y), \pi(y)], \quad  \tilde\pi(x) =
\tilde\pi (x ; \phi(y), \pi(y)]\label{twotwelve} \end{equation}
such that $\widetilde\phi(x)$ and $\tilde\pi(x)$ also satisfy
(\ref{twofour}). (The notation means that $\widetilde\phi$ and $\tilde\pi$
are functions of $x$ but functionals of $\phi(y), \pi(y)$, thus allowing for
non-local field redefinitions.)

Unitary transformations of the fields at one moment of time are an
example of a field redefinition. Under such a transformation
\begin{mathletters}
\label{twosixteen}
\begin{eqnarray}
\tilde\phi(x) & = & U\phi(x)\, U^{-1},
\label{eq:a3}\\
\tilde\pi(x) & = & U\pi(x)\, U^{-1},
\label{eq:b3}
\end{eqnarray}
\end{mathletters}
the fields so redefined satisfy the canonical commutation relations.
\footnote{Indeed, were it not for the possibility of inequivalent
representations
of the commutation relations, (\ref{twosixteen}) would be the most
general field redefinition preserving those commutation relations.}

The operators in a triple $(\{C_\alpha\}, H, \rho)$ may be expressed
as functions either of $(\phi(x),\pi(x))$ or
$(\tilde\phi(x),\tilde\pi(x))$.
In the absence of any external apparatus to give some objective meaning to
the field operators, the description of a triple  in
terms of one set of fields and
momenta is just as valid as the description in terms of another set,
unless we use criteria
such as algorithmic information content.
Such criteria may lead {\it
us} to prefer one description to another but are not intrinsic to quantum
mechanics. The two different descriptions of the same triple --- in terms of
two
different sets of fields and conjugate momenta --- are thus physically
equivalent. Triples of alternative histories, Hamiltonian, and initial
condition may therefore be described in
many different, physically equivalent ways.

Some discussion of the classical analog of this situation may be helpful.
Classically, a fine-grained history is a curve in phase space. The curve
may be described by introducing canonical co\"ordinates $(q^i, p_j)$ on
phase space and giving the functions $\bigl(q^i(t), p_j(t)\bigr)$. However,
any other set of canonical co\"ordinates $(\tilde q^i, \tilde p_j)$,
functions of $(q^i, p_j)$ such that the Poisson bracket ${\{\tilde q^i,
\tilde p_j\}}$ equals $\delta^i_j$, would give equally good and physically
equivalent ways of describing the history. Coarse-grained alternatives can
be constructed using an exhaustive set of mutually exclusive regions of phase
space
analogous to projections. Again these can be described in
many different ways.

\subsection{Equivalent Sets of Histories and Initial Conditions
Represented by Different Operators} \label{sec:f}

Having in hand the above discussion of how the {\it same} operators may be
described in various ways in terms of different quantum fields (or in terms of
various  functions of fields at different times, using the equations of
motion), we now proceed
to describe a notion of physical equivalence between sets of histories and
initial conditions represented by {\it distinct } operators.
{\it Two triples $\bigl(\{C_\alpha\}, H, \rho\bigr)$ and $\bigl(\{\widetilde
C_\alpha\}, \tilde H, \tilde\rho\bigr)$ are physically equivalent if there are
fields and
conjugate momenta $\bigl(\phi (x), \pi (x)\bigr)$ and $\bigl(\tilde \phi
(x), \tilde\pi (x)\bigr)$, respectively, in which the histories,
Hamiltonian, and
initial condition take the same form, for each triple.}

Quantities invariant under field redefinitions are useful in identifying
physically equivalent triples $\bigl(\{C_\alpha\}, H, \rho\bigr)$.
One such quantity is the decoherence functional, which is the same for
two physically equivalent triples
$\bigl(\{C_\alpha\}, H, \rho\bigr)$
and $\bigl(\{\widetilde C_\alpha\}, \tilde H, \tilde\rho\bigr)$. They
both decohere or not to the same accuracy,
and, if they decohere, they have the same probabilities.

Consider an initial density matrix $\rho$ and a set of histories
$\{C_\alpha\}$ made up of sums of chains of projections $\{P^k_{\alpha_k}
(t_k)\}$ at times $t_1,\cdots, t_n$. Let
\begin{mathletters}
\label{twofourteen}
\begin{equation}
\tilde\rho = U\rho\, U^{-1},
\label{aa}
\end{equation}
\begin{equation}
\tilde H = U H U^{-1},
\label{bb}
\end{equation}
and, for each time $t_k$,
\begin{equation}
\widetilde P^k_{\alpha_k} (t_k) = UP^k_{\alpha_k} (t_k) U^{-1},
\label{cc}
\end{equation}
\end{mathletters}
for some unitary fixed transformation $U$, the same for all times $t_k$.
The transformed values of the class operators $\{\widetilde
C_\alpha\}$ are defined by (\ref{twoeight}) with the $P$'s replaced
by the corresponding $\widetilde P$'s.
The operators in the transformed triple
$\bigl(\{\widetilde C_{\alpha}\}, \tilde H,
\tilde\rho\bigr)$ may be regarded either as functions of the fields and
momenta
$(\phi(x),\pi(x))$ or as functions of any other set of fields
and momenta $(\tilde\phi(x), \tilde\pi(x))$ satisfying the
canonical commutation
relations. The same is true for the untransformed  $\bigl(\{
C_{\alpha}\}, H, \rho\bigr)$. The important point is that generally
the operators in the triple
$\bigl(\{\widetilde C_{\alpha}\}, \tilde H, \tilde\rho\bigr)$  will be
{\it different} functions of a {\it given} set of fields and momenta from those
in
$(\{ C_{\alpha}\}, H, \rho\bigr)$. But the former have the same form
in terms of fields $\bigl(\phi(x),
\pi(x)\bigr)$ as the latter
do in terms of the fields $\bigr(\tilde\phi (x), \tilde \pi (x) \bigr)$,
where $\bigr(\tilde\phi (x), \tilde \pi (x) \bigr)$
are defined by (\ref{twosixteen}) and obey the same
canonical commutation relations as the  $\bigl(\phi(x),
\pi(x)\bigr)$. Moreover, the decoherence functional is the same for
the old triple and the new. Thus, the two triples are physically equivalent
(in the sense
defined above) and we propose that they be identified with each other.

The
analogous classical situation may be helpful in understanding the notions of
physical equivalence and identification that we have introduced. As we
mentioned earlier, the
classical analog of a fine-grained history is a curve in phase space and the
analog of a projection is a region of phase space. The analog of an
initial density matrix is an initial  phase space distribution. A canonical
transformation may be used to transform these into new curves, new regions,
and new initial distributions. However, for a closed system, in the
absence of an imposed labeling by physical quantities of the points in
phase space, the new triple
of histories, Hamiltonian,  and initial
condition is physically indistinguishable from the old triple
because it has the same description in terms of the canonically
transformed co\"ordinates and momenta that the old one did in terms of
the original co\"ordinates and momenta. Again we can identify the two
triples. Instead of distinguishing triples we then distinguish
equivalence classes of triples. Of course, those theorists who prefer
to impose --- even
for a closed system --- a labeling by physical quantities of the rays
in Hilbert space or the points in phase space (as discussed in Section
D) have, by that convention, selected one member of each equivalence
class.

\subsection{Unitary Transformations That Leave the Initial Density Matrix
Fixed}
\label{sec:g}

Of all the unitary transformations (\ref{twofourteen}) yielding physically
equivalent sets of histories and initial conditions, the ones that leave
the initial density matrix fixed \begin{equation} \rho = U\rho U^{-1}
\label{twonineteen}
\end{equation}
are of special importance for the problem of quasiclassical realms. If the
initial density matrix is pure or close to pure, there will be a great
variety of
such $U$'s because out of the Hilbert space of the universe
only one vector or a small subset of vectors needs to be
left fixed.

Suppose we have an initial $\rho$ and seek to compute the sets of histories
representing any quasiclassical realms that emerge from $\rho$ and
$H$. We might be
tempted to think that if we found one sequence of sets of $P$'s
representing a quasiclassical realm we could find many others simply by
acting on the $P$'s with a $U$ that preserves $\rho$. However, all
those sets are physically
equivalent and identified with one another. They represent the same
quasiclassical realm. The universe may also exhibit essentially inequivalent
quasiclassical realms, but they are not to be found in this manner ---
simply by redefining fields. Any measure of classicality
should be defined on equivalence classes of physically equivalent
histories.

\subsection{The Information Content of a Physical Equivalence Class}
\label{sec:h}

Having pointed out some transformations that leave unchanged a set of
decohering histories such as a quasiclassical realm, we should now discuss
where the information actually lies that characterizes such a decohering set.
Consider, for simplicity, the case of a pure $\rho=|\Psi\rangle\langle
\Psi|$. Then, what it means for a set of histories $\{C_\alpha\}$
to decohere exactly is that all non-vanishing vectors $C_\alpha|\Psi
\rangle$ are orthogonal to one another, with their norms giving the
probabilities of the alternative histories labeled by the index $\alpha$.
Since the histories are exhaustive, we have $|\Psi\rangle = \Sigma_\alpha
C_\alpha
|\Psi\rangle$. The state vector $|\Psi\rangle$ is thus resolved into
components, corresponding to branches, in a basis consisting of the
non-vanishing
vectors $C_\alpha|\Psi\rangle$ (normalized to unity) and any other
set of unit vectors, orthogonal to one another and to all the  non-vanishing
 $C_\alpha|\Psi \rangle$, that make the basis complete.

Since, however, sets of branches
$C_\alpha|\Psi\rangle$  that are related
by any fixed unitary transformation $U$ are physically equivalent, the only
physical information
contained in the relations among the vectors is that a normalized
state $|\Psi\rangle$ is resolved into
a set of components  (labeled by those $\alpha$'s
such that $C_\alpha|\Psi\rangle \neq 0$)
with particular norms, which are the probabilities, and that zero probability
is assigned to the basis vectors that are orthogonal to all the remaining
non-vanishing $|C_\alpha\rangle$. Besides the list of norms (probabilities)
and zeros, there is no invariant information in the relation among the vectors.
{\it What does carry information, other than just probabilities, is the
explicit narrative content of the $\{C_\alpha\}$ --- sums of chains
of projections expressed in terms of field operators --- compared with
the form of the initial condition
$|\Psi\rangle$ expressed in terms of the same field operators.}

In that connection, it is interesting to remark that at one time
\cite{GH90b} we introduced an entirely different set of equivalence classes
from the ones
discussed in this article. In the earlier work, two sets of histories
$\{C_\alpha\}$
and $\{C'_\alpha\}$ are treated as equivalent if $C_\alpha|\Psi\rangle
=C'_\alpha|\Psi\rangle$ for every $\alpha$. Incorporating the results
obtained here,
we see that the list of probabilities of histories is the only invariant
property
of an equivalence class of the type
we defined then. All the other properties of histories relate to
variation within one of those equivalence classes, that is, variation
of the different
operators $\{C_\alpha\}$, with their different narratives, leading to the
same resolution of the state vector $|\Psi\rangle$ of the universe into
orthogonal  branches $\{C_\alpha|\Psi
\rangle\}$. That resolution, together with the content of the
$|\Psi\rangle$ and of the $\{C_\alpha\}$ --- both  expressed in terms of
a given language of
field operators --- is the basis of the interpretation of quantum mechanics,
at least if $\rho$ is pure.

\subsection{The Approximate Quantum Mechanics of Measured Subsystems}

\label{sec:i}

Measurement situations are most accurately described in the quantum
mechanics of a closed system that contains both measuring apparatus and
measured subsystem. However, those situations can also be treated
to an excellent
approximation by the approximate quantum mechanics of measured
subsystems (AQMMS) (aka the ``Copenhagen'' formulation),  which is
so familiar from textbooks.
This Section discusses the connection between the notions of physical
equivalence that hold in these two formulations.

In the approximate quantum mechanics of measured subsystems, the joint
probability of a sequence of ``ideal'' measurements carried out on a
subsystem with a pure (for simplicity) initial state $|\psi\rangle$ is
\begin{equation}
p\left(\alpha_n, \cdots, \alpha_1\right) = \left\| s^n_{\alpha_n} (t_n)
\cdots\cdots s^1_{\alpha_1} (t_1) |\psi\rangle \right\|^2\ . \label{twoone}
\end{equation}
Here, for a given value of $k$, the set $\{s^k_{\alpha_k} (t_k)\}$
consists of projection operators (in the Heisenberg picture)
representing the possible outcomes, enumerated by the index $\alpha_k$,
of the measurement carried out at time $t_k$. Thus, if the
subsystem consisted of a single particle and the measurement at time $t_k$
localized the particle to one of a set of position intervals
$\Delta^k_{\alpha_k}, \alpha_k = 1, 2, \cdots$,  then the operators
$\{s^k_{\alpha_k} (t_k)\}$ would be projections onto those intervals at
time $t_k$. State vectors, projections, etc. in (\ref{twoone}) all refer to
the Hilbert space ${\cal H}_s$
of the measured {\it sub}system.

The physical consequences of AQMMS are left unchanged by a trivial
relabeling of the Hilbert space ${\cal H}_s$ of the kind described in
Section B for closed systems.  Such a relabeling is implemented by an
alias (passive)
unitary transformation of all operators and vectors.
However, the notions of physical equivalence of the kinds discussed in
Sections C and F have a different character for AQMMS.

The probabilities (\ref{twoone}) are unchanged by a reassignment of the
times to the sequences of measurements as long as the operators
representing those measurements are unchanged. However, in AQMMS
one presumes that there are clocks external to the subsystem
that give a physical meaning to time, so that sets of projection
operators on the Hilbert space of the subsystem that are
assigned to different times correspond to physically distinct
alternatives. Specifically, they correspond to measurements on
the subsystem carried out at different times as determined by
the external clock. Reassignment of the times, therefore, does not
lead to a physically equivalent set of histories in AQMMS --- in contrast to
the quantum mechanics of closed systems, where
there are no external clocks. (Of course, even for a closed system one
could arbitrarily specify time labels and thus remove the freedom to
reassign the times.)

Next, consider a unitary transformation $u$ of the kind discussed in Section
F, acting only on the Hilbert space ${\cal H}_s$ of the measured
subsystem.
The values of the probabilities (\ref{twoone})
are left unchanged by the substitutions \begin{mathletters}
\label{eq:twotwo}
\begin{eqnarray}
|\psi\rangle & \to & |\widetilde\psi\rangle = u|\psi\rangle , \label{eq:aa}\\
s^k_{\alpha_k} (t_k) & \to & \tilde s^k_{\alpha_k} (t_k) = u
s^k_{\alpha_k} (t_k) u^{-1}, \label{eq:bb}
\end{eqnarray}
\end{mathletters}
where $u$ is the same for all $t_k$.
However, in AQMMS, different sets of
orthogonal projections $\{s^k_{\alpha_k}(t_k)\}$ are presumed to describe
the alternative outcomes of distinct measurements, with distinct kinds of
apparatus.
Given a set of projections, it is in principle possible to construct an
apparatus that measures the represented
alternatives and distinguishes them from those represented by any other set
of projections  \cite{Lamxx}. The
measurements represented by the $\{s^k_{\alpha_k}(t_k)\}$, and the
$\{\tilde s^k_{\alpha_k}(t_k)\}$ are different, despite the fact that they have
the same probabilities, unless, of course, $u$ commutes with {\it all} the
$\{s^k_{\alpha_k}(t_k)\}$, so that the set $\{s^k_{\alpha_k}(t_k)\}$
is the same as the set $\{\tilde s^k_{\alpha_k}(t_k)\}$. The only
unitary operators that commute with all projections are multiples of the
identity \begin{equation} u=e^{i\alpha} I,
\label{twothree}
\end{equation}
or, when there are superselection rules, multiples of the identity
with different phases on  different superselection sectors.
Thus, $|\psi\rangle$
is physically equivalent to $e^{i\alpha}|\psi\rangle$, or, in other words,
physical states in quantum mechanics are represented
by {\it rays} in Hilbert space.\footnote{See
\cite{BS93} for an insightful discussion of rays.}

AQMMS is an approximation to the more general quantum mechanics of
closed systems. Examination of models of measurement situations to which
this approximation applies shows how the more restrictive notions of
physical equivalence of AQMMS emerge in the more general context.

In a standard kind of closed system model of a measurement situation
\cite{Har91a}, the Hilbert space $\cal H$ is assumed factored into two
parts: a Hilbert space ${\cal H}_s$ representing the measured subsystem
and a Hilbert space ${\cal H}_r$ representing the rest, including the
measuring apparatus. It is convenient to think of separate canonical
pairs of co\"ordinates and momenta $z^a=\{y^i,p_i\}$ that act on
${\cal H}_s$, and others $Z^A=\{Y^I, P_I\}$ that act on
${\cal H}_r$. The interaction between the two subsystems is described in
by an interaction Hamiltonian $H_{\rm int}$ that is a function of both
kinds of variables as well as the time.  Appropriate kinds of initial
states in which the
variables of ${\cal H}_r$ and ${\cal H}_s$ are uncorrelated evolve under
the action of the total Hamiltonian containing $H_{\rm int}$ into states
in which the value of some physical quantity $f(z,t)$ on ${\cal H}_s$ at
time $t$
becomes tightly correlated with the value of a physical quantity $F(Z,T)$
on ${\cal H}_r$ at a possibly distinct time $T$. In that way $f(z,t)$ is
``measured'' by the subsystem
represented by ${\cal H}_r$. Under suitably idealized conditions
eq.(\ref{twoone}) --- representing the
approximate quantum mechanics of measured subsystems --- approximates the
probabilities of the outcomes of successions of such measurements.

The connection between the notions of physical equivalence in AQMMS and
in the closed system measurement model adumbrated above can be understood
in the following way: We mentioned earlier that subspaces of Hilbert space
could
be labeled by requiring some physical quantities to have particular
functional forms. Suppose that this is done in such a way as to label
the subspaces of ${\cal H}_r$ and further to specify the form of the
interaction Hamiltonian $H_{\rm int}(z,Z,t)$. The rules of AQMMS are not
affected by such a choice since they refer neither to ${\cal H}_r$ nor
to $H_{\rm int}$.

In the closed system, two triples $( \{C_\alpha\}, H, \rho)$ and
$(\{\tilde C_\alpha\}, \tilde H, \tilde\rho)$ are physically equivalent
if the operators in the two triples are related by a constant unitary
transformation or by reassignment of the times. However, except in
special cases, such a transformation or reassignment
can be expected to change the form of operators on ${\cal
H}_r$ or the form of $H_{\rm int}(z,Z,t)$, with the exception of trivial
unitary transformations that are multiples of the identity on each
subspace. For instance, unitary transformations of the form $u \otimes I$,
with $u$ unitary on ${\cal H}_s$, would not affect the form of the operators
on ${\cal H}_r$, but would generally change the form of $H_{\rm
int}(z,Z,t)$. Thus the choice of the functional form of $H_{\rm int}$ and
of a sufficient number of operators on ${\cal H}_r$ fixes an essentially
unique representative of the physical equivalence class of descriptions
of the measurement model. The remaining freedom will generally consist
of unitary transformations that can be written $e^{i\alpha}I_s \otimes
e^{i\beta}I_r$. We see that when AQMMS is viewed as an approximation to the
quantum mechanics of closed systems with a fixed form of $H_{\rm int}$
and a physical labeling of the subspaces of ${\cal H}_r$, we recover the
appropriate notion of physical equivalence for AQMMS that was described
earlier in this Section.

\section{IGUSes and Quasiclassical Realms}\label{sec:III}

As we mentioned in the Introduction, sets of histories constituting
quasiclassical realms are important in quantum mechanics because they
are utilized by IGUSes. We human observers, for instance, most often
describe the world about us using coarse-grained histories that
distinguish ranges of values of familiar quantities of classical physics.
In this Section, we discuss, not completely, but
in more detail than we have before, how IGUSes are characterized and
how the probabilities of their existing and behaving in certain ways
are in principle predictable from quantum cosmology.

Human beings, bacteria, and computers equipped with certain kinds of
hardware and/or software are all examples of IGUSes at various levels
of complexity. IGUS is our name for a complex adaptive system in the
context of quantum mechanics. Roughly, an IGUS is a subsystem of the
universe that
makes observations and thus acquires information,
makes predictions on the basis of that information using some approximation
(typically very crude) to the true quantum-mechanical laws of nature, and
exhibits behavior based on those predictions.
In general a complex adaptive system has following features\footnote{For
a more complete discussion of complex adaptive systems, see
 \cite{GM94}, \cite{GMxx}, \cite{GMyy}}:
(1) It identifies and records regularities in an input data stream.
(2) It compresses these regularities into a schema, which can be thought
of as a model or theory.
(There are typically variant schemata
in competition with one another.)
(3) A schema, enriched by further data, is used for describing the world, for
predicting the future, and for prescribing
behavior of the complex adaptive system as well as regulating the acquisition
of further information.
(4) These interactions with the world give rise to selection pressures
exerted back on the competition of schemata, resulting in evolutionary
adaptation.

All known complex adaptive systems on Earth are related in some
way to life. They range from the prebiotic chemical reactions that
produced life, through biological evolution, the functioning of
individual organisms and ecological systems, the process of thinking
in humans (and other animals), and the operation of mammalian immune
systems, to the functioning of computers programmed to evolve
strategies for playing games. All of them utilize the usual quasiclassical
realm. That is, the
input data stream  and the
consequences of a schema in the world are all describable in essentially
classical terms
by ranges of values of usual quasiclassical operators. We might call such
IGUSes
entirely usual quasiclassical IGUSes (EUQUIGUSes).

Sets of alternative histories for EUQUIGUSes
are necessarily coarse
grainings of the usual quasiclassical realm. The predictions of quantum
cosmology for
the individual histories in these coarse-grained sets are the
probabilities of those histories.
For example, the histories of the universe associated with the usual
quasiclassical realm might be
partitioned into those that exhibit IGUSes at certain times
and places and those that
never do. In this way the probability of existence of
EUQUIGUSes becomes in principle a calculable question in
quantum cosmology. The histories of the usual quasiclassical realm
might be partitioned according to different evolutionary tracks of classes
of IGUSes; in this way their evolution could be discussed.
For example, by using an appropriate coarse graining,
one might ask whether IGUSes evolve preferentially
near type G stars. One could
in principle calculate the conditional probabilities for alternative
behaviors of an individual IGUS given its input data, or
for the alternative evolutionary tracks of species given different
selection pressures. (Such conditional probabilities, while depending
on the initial condition of the universe, may be especially sensitive to
the information about the specific past history
that sets the conditions.) We should not pretend that it is practical
to calculate the probabilities of histories such
as we are discussing. However, it is in this manner
that the nature, behavior, and evolution of
EUQUIGUSes would be predictable in principle,
from a fundamental theory of the elementary particles and the
initial condition of the universe, in the form of probabilities
arising from the quantum mechanics of closed systems.

\section{Other Realms}\label{sec:IV}

Alternative histories referring to IGUSes need not be restricted to
coarse grainings of the usual quasiclassical realm as they were in
the preceding Section.
By relaxing the assumption that all features of an IGUS are describable
in usual quasiclassical terms, we may investigate a broader class of questions
pertaining
to IGUSes. Such questions involve decohering sets of
alternative coarse-grained histories of the universe defined by
operators
other than those of the usual quasiclassical realm. We can say that they are
defined by realms other than the usual quasiclassical realm.

As we have emphasized before \cite{GH90a}, quantum theory itself does not
discriminate between different sets of alternative decohering histories
of a closed system (different realms), except by measures of their
coarse graining, classicality, {\it etc.} As stressed by
Griffiths \cite{Gri84}, Omn\`es \cite{Omnsum}, and -- more recently --
by Dowker and Kent \cite{DKup}, great care is therefore needed in the
use of ordinary language in dealing with quantum mechanics. In particular,
different language should be used for discussions of the properties of
a {\it single}
realm from that used to discuss the relationships between {\it different}
realms. We recommend in particular that words like ``exist'', ``happen'',
``occur'' {\it etc.} should be used only to refer to alternatives
within a single realm, or else to projections that are perfectly
correlated with such alternatives, as when a quantum-mechanical
operator is measured by a classical apparatus.  That way  these words would
have meaning in terms of
quantum-mechanical probabilities, as expected. For example, we have
discussed elsewhere
\cite{GH90a,Har91a}, what is meant by an event having ``happened''
in the past. In a decohering set of histories describing
certain present data as well as alternatives in the past that include
the event, the conditional probability for the occurrence of the event in
the past --- given the present
data --- is near unity, while the conditional probability for alternatives
to the event is near zero.
When discussing {\it different} realms
as features of the theory, we recommend not using words (such as``exist'',
``happen'', or ``occur'' ) that could be expected to have a probabilistic
meaning. The reason is that quantum mechanics does not assign probabilities to
different realms. Rather we suggest using phrases like ``the theory
exhibits a realm with this or that property'' or the theory ``allows
a realm....''. Adhering to these usages will be especially important
for clarity in the discussions in the remainder of this Section of IGUSes in
realms different from the
usual quasiclassical one
and for those of the relationships between realms in Sections  V.

The four defining properties of an IGUS introduced in the previous
Section may be applicable to more general realms than the usual
quasiclassical one considered there.
With such a definition, the probability of occurrence of IGUSes --- and their
nature, evolution, and behavior ---
could be investigated in realms very different from the usual
quasiclassical one.
Input data, selection pressures, etc would be described in terms
of correspondingly different alternatives. For those realms in which
one can meaningfully distinguish the branches on which
IGUSes evolve from those on which they have not evolved, one may ask for the
total probability of the branches that do exhibit IGUSes.
This probability for IGUSes may then be compared between different
realms. If the probability is high mainly for quasiclassical
realms or coarse grainings of such realms, that is one way to give a meaning to
the
to the conjecture that IGUSes evolve
primarily on branches of quasiclassical realms.\footnote{ The fact that
known IGUSes have evolved to
utilize mostly quasiclassical alternatives in their input data stream
provides an answer to  questions such as:
``If the universe is in a superposition of quasiclassically distinct
histories, why isn't it seen   in a superposition?'' That sort of
question
is particularly relevant for cosmology .
A postulated pure initial quantum state of the universe is typically a
superposition of quasiclassical branches with differing
positions of individual stars    (among many other things) at a given
time.
The eyes of birds, for example, have evolved to distinguish such
quasiclassical alternatives, rather than
to discriminate between alternative superpositions of quasiclassical
branches.
In each branch, certain registrations of what birds see are correlated with
fairly definite positions of bright stars. Thus they detect
stars  in particular places in the sky at a given time (and sometimes use them
for navigation)
even though the universe may be said to be in a
superposition of branches in which the stars have various positions at
that time --- in
the sense that its initial Heisenberg state vector is a sum of the
corresponding
branch state vectors.
}
Even if a high probability for IGUSes is not restricted to
quasiclassical realms, one might establish the requirements for IGUSes
by comparing different realms. For example, one could investigate the
level of determinism necessary for IGUSes by comparing the probabilities
for their evolution in realms exhibiting varying levels of determinism,
that is, by comparing the probability of IGUSes in the usual quasiclassical
realm
with that in realms that are much less deterministic or
much more deterministic.

The representation of the universe as a quantum computer, as discussed
by Seth Lloyd \cite{Llo93} and in earlier work cited by him, comes close
to exhibiting a realm totally different from the usual quasiclassical
one but much more deterministic.
The coarse graining consists entirely of restriction to equally
spaced instants of time. In a particular basis in Hilbert space, the
different vectors represent different initial conditions of a computer
consisting of the entire universe.  From  any of these initial states,
the universe
proceeds deterministically from one state to another
as the equal intervals of time ``tick'' by.
The equations of motion here are discrete and exactly deterministic
instead of continuous and approximately deterministic, but there
is  some resemblance to a quasiclassical realm  despite
the absence of branching after the first tick.
Presumably computer-based IGUSes are to be found with probability near
one within this representation
of the universe.

Apart from the somewhat artificial example of the universe as a quantum
computer, there is the possibility that the quantum mechanics of the universe
as a consequence of the
initial condition and Hamiltonian could exhibit a realm that
is essentially different from the usual quasiclassical one but
characterized by a high measure of classicality.
Such a distinct quasiclassical realm would
be a set of alternative histories, obeying a realistic principle of
decoherence, displaying with high probability patterns of deterministic
correlation described by effective equations of motion,
and maximally refined subject to those conditions. It would
differ from the usual quasiclassical realm because its alternatives would
not be describable (even by means of a unitary transformation
preserving $\rho$) in terms of ranges of values of integrals over
small spatial volumes of the familiar field and
density operators of classical physics, but would have to be described
in terms of other operators instead. Thus the approximate and
phenomenological deterministic laws would be different from those of the
usual classical physics.\footnote{ A. A. Starobinsky tells us that such
distinct quasiclassical realms are called ``goblin worlds" by some science
fiction writers. Using terminology such that ``realm" refers to a set of
alternative histories, we are discussing goblin realms.}
Although in pointing out some features of
the usual quasiclassical operators that crudely characterize them,
we raised
the suspicion that the quasiclassical realm of everyday experience might
be essentially unique \cite{GH90a}, it seems possible that the universe
exhibits
truly distinct quasiclassical realms.

If a quasiclassical realm different from the usual one is exhibited
then there is no reason to suppose that
it might not possess coarse grainings describing  the
evolution and
behavior of complex adaptive systems acquiring and utilizing
information. That is, the universe might exhibit IGUSes in distinct
quasiclassical realms, and the laws of quantum mechanics would not prefer
one to another. Quantum theory supplies in principle the probability of
such IGUSes in each realm, although those probabilities are far beyond
our ability to compute in practice.
In the next Section we discuss possible relationships between such
realms.

\section{Relations Between Realms}\label{sec:V}
If different realms exhibit IGUSes, we may investigate certain
relations between them. Probabilistic predictions
concerning the relationships between IGUSes in two different realms
may be made by using a decohering set of histories containing
alternatives referring to IGUSes in one realm and also alternatives
referring to IGUSes in the other realm, provided the decoherence of the hybrid
set follows from the initial condition and Hamiltonian.
The problem of drawing inferences in one realm concerning IGUSes using a
distinct realm is then not so very different from that involved in
ordinary searches for extraterrestrial intelligence. There, we observe
projections
accessible to us and try to infer from their particular values that they
are signals from IGUSes, say because they can be seen to be an encoding of
$\pi$ or of the periodic table of chemical elements. Such discussions
are most often carried out using the usual quasiclassical realm.
However, one could also use a realm containing
alternatives from the usual quasiclassical realm that
described our observations, and in addition alternatives from the other
realm describing other IGUSes. Using such a hybrid realm, one could
compute probabilities for alternatives referring to those other IGUSes.
In the following we describe some kinds of hybrid realms.

The simplest example refers to alternatives describing IGUSes in
one realm that are highly correlated with histories consituting a coarse
graining of another realm. This is very like an ordinary measurement
situation
in which a value of a non-quasiclassical operator such as a spin
becomes almost exactly correlated with a range of values of a usual
quasiclassical one.
Then IGUSes making use of one realm
could conceivably draw inferences about IGUSes in another by seeking or
creating
``measurement situations''
in which an alternative of one realm is correlated almost
perfectly  with an alternative from the other.

Even in the absence of nearly exact correlations, it is possible that
probabilistic inferences about some features of IGUSes evolving in 
different realms
could be drawn making use of this kind of hybrid realm with some alternatives
drawn from both.  

To give a very simple example of how questions may be asked in this way
about IGUSes
using realms different from the usual quasiclassical one, we may
investigate the probabilities that IGUSes, otherwise described in
quasiclassical terms,  evolve to use an input data stream containing,
in part, alternative ranges of values of operators essentially different from
the usual quasiclassical ones. IGUSes like ourselves that measure the
values of highly quantum-mechanical variables are examples. Sets of
histories describing such alternative evolutionary tracks contain
mostly operators of the usual quasiclassical kind, but occasionally
non-quasiclassical operators describing measured alternatives,
which are highly correlated with certain of the usual
quasiclassical ones.
A low probability that IGUSes evolve to make
direct use of non-quasiclassical alternatives is another way in which our
conjecture that IGUSes evolve to exploit the regularities of a
quasiclassical realm can be given a meaning within quantum mechanics.

No particular IGUSes, such as human beings, play any distinguished
role in the formulation of quantum mechanics that we are using.
If the universe exhibits IGUSes in realms essentially different from the
usual quasiclassical one, that does
not constitute a paradox, but rather an
intriguing example of the richness of possibilities that may be shown
by a quantum universe.

\acknowledgements
We thank H.F.~Dowker, R.~Griffiths, J.~Halliwell, A.P.A.~Kent, and D.
Page for critical
readings of the manuscript.
One of the the authors (JBH) thanks R. Penrose for a stimulating conversation.
The authors are grateful for the hospitality of the Aspen Center for
Physics and of
the Isaac Newton Institute
for Mathematical Sciences, University of Cambridge, UK.
The work of MG-M was supported by grants to the Santa Fe Institute from
J. Epstein and G. Gartner. The work of JBH was supported in part by
NSF grant PHY90-08502.


\begin{references}

\bibitem{GH90a} M.~Gell-Mann and J.B.~Hartle in {\sl Complexity, Entropy,
and the Physics of Information, Santa Fe Institute Studies in the Sciences
of Complexity}, Vol.
VIII, ed. by W. \.Zurek, Addison Wesley, Reading (1990) or in {\sl
Proceedings of
the 3rd
International Symposium on the Foundations of Quantum Mechanics in the Light of
New Technology} ed.~by S.~Kobayashi, H.~Ezawa, Y.~Murayama, and S.~Nomura,
Physical Society of Japan, Tokyo (1990).

\bibitem{DKup} H.F.~Dowker and A.~Kent (to be published).

\bibitem{GH93a} M.~Gell-Mann and J.B.~Hartle, \journal Phys. Rev. D, 47,
3345, 1993

\bibitem{PZ93}J.P. Paz and W.H. \.Zurek \journal Phys.Rev.D, 48, 2728, 1993.

\bibitem{GH95} M.~Gell-Mann andd J.B.~Hartle (to be published).

\bibitem{Har95} J.B.~Hartle, {\sl Quasiclassical Domains in a Quantum
Universe}, in {\sl Proceedings of the Lanczos Centenary
Conference},
North Carolina State University, December, 1992, ed. by J.D. Brown,
M.T.~Chu, D.C.~Ellison, and R.J.~Plemmons, SIAM, Phildelphia, (1994).


\bibitem{Llo93} S.~Lloyd, private communication and \journal Phys. Rev.
Lett., 71, 943, 1993.


\bibitem{Har91a} J.B.~Hartle, {\sl The Quantum Mechanics of Cosmology}, in
{\sl Quantum Cosmology and Baby Universes: Proceedings of the 1989
Jerusalem Winter School for Theoretical Physics}, ed. by S.~Coleman,
J.B.~Hartle, T.~Piran, and S.~Weinberg, World Scientific, Singapore
(1991).


\bibitem{Lamxx} W.~Lamb, \journal Physics Today, 22, 23, 1969.

\bibitem{BS93} J.C.~Solem and L.C.~Biedenharn, \journal Found. Phys., 23,
185, 1993.

\bibitem{Harpp} J.B.~Hartle, {\it Spacetime Quantum Mechanics and the
Quantum Mechanics of Spacetime} in {\sl Gravitation and Quantizations:
Proceedings of the 1992 Les Houches
Summer School}, ed. by B. Julia and J. Zinn-Justin, North Holland
Publishing Co, Amsterdam, (1994), grqc/9304006.


\bibitem{Har91b} J.B.~Hartle, \journal Phys. Rev., D44, 3173, 1991.

\bibitem{Gri84} R.~Griffiths, \journal J. Stat. Phys., 36, 219, 1984.

\bibitem{GH93b} M.~Gell-Mann and J.B.~Hartle, in {\sl Proceedings of the
NATO Workshop on the Physical Origins of Time Asymmetry, Mazag\'on, Spain,
September 30-October 4, 1991} ed. by J. Halliwell, J. P\'erez-Mercader, and
W. \.Zurek, Cambridge University Press, Cambridge (1994), grqc/9304023.

\bibitem{Omnsum} R.~Omn\`es, \journal J. Stat. Phys., 53, 893, 1988,
\journal ibid, 53, 933, 1988; \journal ibid, 53, 957, 1988; \journal ibid,
57, 357, 1989; \journal
Rev.~Mod.~Phys., 64, 339, 1992.


\bibitem{GH90b} M. Gell-Mann and J.B. Hartle in the {\sl Proceedings of
the 25th International Conference on High Energy Physics, Singapore, August
2-8, 1990},
ed.~by K.K.~Phua and Y.~Yamaguchi (South East Asia Theoretical Physics
Association
and Physical Society of Japan) distributed by World Scientific, Singapore
(1990).

\bibitem{GM94} M. Gell-Mann, {\sl The Quark and the Jaguar}, W. H. Freeman,
New York (1994).

\bibitem{GMxx} M.~Gell-Mann,  {\sl  Complexity and Complex
Adaptive Systems}, in {\sl The Evolution of Human Languages}, Santa Fe
Institute Studies in
the Sciences of Complexity, Proc. Vol. X, ed. by
M.~Gell-Mann and J.A.~Hawkins, Addison-Wesley, Reading, MA (1992).

\bibitem{GMyy} M.~Gell-Mann, {\sl Complex Adaptive Systems} in {\sl
Complexity: Metaphors, Models, and Reality}, Santa Fe Institute Studies in
the Sciences of Complexity,
Proc. Vol. XIX, ed. by G.A.~Cowan,  D.~Pines, and D.~Meltzer,
Addison-Wesley,
Reading, MA, (1994).



\end{references}
\end{document}